\documentclass{eas}
\usepackage{graphicx}
\begin{document}

\title{GTC/OSIRIS observations of RWT 152, a case study of a planetary nebula with an sdO central star} 
\author{A. Aller}\address{Departamento de Astrof\'isica, Centro de Astrobiolog\'ia - Spanish Virtual Observatory}\secondaddress{Departamento de F\'isica Aplicada, Universidade de Vigo}
\author{L. F. Miranda}\address{Instituto de Astrof\'isica de Andaluc\'ia-CSIC}
\author{L. Olgu\'in}\address{Departamento de Investigaci\'on en F\'isica, Universidad de Sonora}
\author{E. Solano}\sameaddress{1}
\author{A. Ulla}\sameaddress{2}

\begin{abstract}
RWT\,152 is one of the few planetary nebula with an sdO central star. We present subarcsecond red tunable filter imaging and intermediate-resolution, long-slit spectroscopy of RWT\,152, obtained with OSIRIS/GTC, which allow us to describe in detail its morphology and to obtain its physical conditions and chemical abundances. 
\end{abstract}
\maketitle

\section{Introduction}
Hot subdwarf O stars (sdOs) represent a small fraction of the central stars of planetary nebulae (PNe) known so far. They are blue and evolved objects in their way to the white dwarf cooling sequence. Their origin is still controversial and several evolutionary paths, e.g. post-AGB evolution, can lead sdOs to their current position in the HR diagram.
  However, it is surprising that only about 18 PN+sdO systems are known to date (Aller et al. 2015), a low number as compared with the large fraction of sdOs without associated PNe ($\geq$800, \O{}stensen 2006). Studying in detail the properties of these objects (morphology, chemical abundances, etc.) is desirable in order to constrain their formation and evolutionary history. 

\section{Observations and results}
Images and intermediate-resolution, long-slit spectra were obtained with OSIRIS at the Gran Telescopio Canarias (GTC), on the Observatorio Roque de los Muchachos (La Palma, Canary Islands). For the imaging, the red tunable filter (RTF, Cepa et al. 2003) was used, with a spectral range of 6490-6600\AA. 
Figure\,1 displays the false-color H$\alpha$ image of RWT\,152 after the reduction process. It shows a bipolar shell surrounded by a circular halo. The two main lobes, slightly different to each other, appear composed of multiple small bubbles. A bright region traces the equatorial plane of the shell, which most probably corresponds to the ring-like structure identified in high-resolution spectra (Aller et al. 2015). The main shell and the central star are displaced with respect to the centre of the halo suggesting interaction of the halo with the ISM.  

For the spectra, the gratings R2500U, R2500V, R2500R and R2500I were used. They reveal very weak [Ne\,{\sc iii}], [O\,{\sc iii}], [Ar\,{\sc iii}], [S\,{\sc iii}], He\,{\sc i} and H emission lines. The lack of He\,{\sc ii}$\lambda$4686 in emission and the line intensity ratio [O\,{\sc iii}]/H$\beta$ $\simeq$ 8 indicate a very low-excitation PN. However, the absence of [S\,{\sc ii}], [O\,{\sc i}], [O\,{\sc ii}], and [N\,{\sc i}] emission lines is very peculiar for this low excitation, suggesting a density-bounded PN. The chemical abundances derived and the high peculiar velocity ($\sim$ 92-131 km\,s$^{-1}$) obtained from its heliocentric systemic velocity (see Aller et al. 2015) strongly suggest that RWT 152 is a type III or IV PN.

     \begin{figure}
  \centering
  \includegraphics[width=0.65\textwidth]{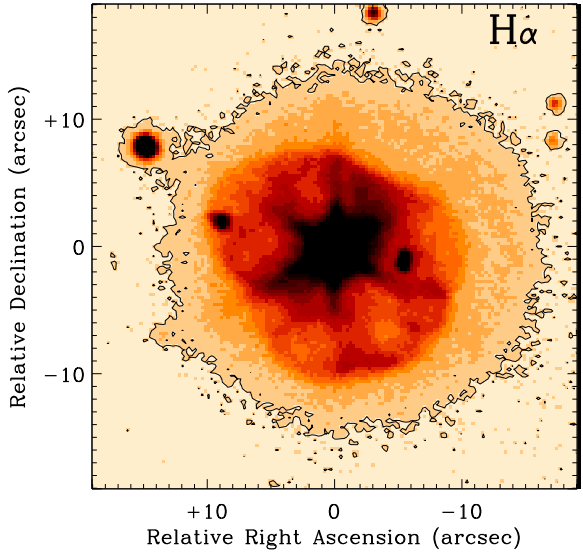}
      \caption[]{False-color H$\alpha$ image of RWT\,152. The contour traces the halo at the 3$\sigma$ level above the background. The spatial resolution is 0.7 arcsec.}
      \end{figure}

{\it Acknowledgements}: Grants AYA 2011-24052, and AYA 2011-30228-C3-01 from the Spanish MINECO, and grant INCITE09312191PR from the Xunta de Galicia, all partially funded by FEDER funds; and grant 12VI20 from University of Vigo.


\end{document}